\documentclass[pof,reprint]{revtex4-1}

\usepackage{graphicx}

\newcommand{\eg}{{\bf e}}
\newcommand{\xg}{{\bf x}}

\newcommand{\deltag}{\mbox{\boldmath $\delta$}}
\newcommand{\mdel}{\mbox{\boldmath $\delta$}}

\newcommand{\etag}{\mbox{\boldmath $\eta$}}
\newcommand{\xig}{\mbox{\boldmath $\xi$}}
\newcommand{\1}{\mbox{\boldmath $1$}}

\newcommand{\fm}{\langle f \rangle}

\newcommand{\uu}{\langle u \rangle}
\newcommand{\ui}{\langle u_i \rangle}
\newcommand{\uj}{\langle u_j \rangle}

\newcommand{\p}{\langle p \rangle}

\newcommand{\dR}{I \! \! R}

\newcommand{\mcalC}{{\cal C}}

\begin{document}

\title{Variable scale filtered Navier-Stokes Equations. A new procedure to deal with the associated commutation error}
\author{M. Iovieno}
\affiliation{%
Doctorate Program in Fluid Dynamics at the Politecnico di Torino, Corso Duca degli Abruzzi 24, 10129 Torino, Italy.}%
\author{D. Tordella}
\affiliation{%
Dipartimento di Ingegneria Aeronautica e Spaziale, Politecnico di Torino, Corso Duca degli Abruzzi 24, 10129 Torino, Italy.}%

\begin{abstract}
A simple procedure to approximate the noncommutation terms that arise whenever it is necessary to
use a variable scale filtering of the motion equations and to compensate directly the flow solutions
from the commutation error is here presented. Such a situation usually concerns large eddy
simulation of nonhomogeneous turbulent flows. The noncommutation of the average and
differentiation operations leads to nonhomogeneous terms in the motion equations, that act as source
terms of intensity which depend on the gradient of the filter scale $\delta$ and which, if neglected, induce
a systematic error throughout the solution. Here the different noncommutation terms of the motion
equation are determined as functions of the $\delta$ gradient and of the $\delta$ derivatives of the filtered
variables. It is shown here that approximated noncommutation terms of the fourth order of accuracy,
with respect to the filtering scale, can be obtained using series expansions in the filter width of
approximations based on finite differences and introducing successive levels of filtering, which
makes it suitable to use in conjunction with dynamic or mixed subgrid models. The procedure
operates in a way which is independent of the type of filter in use and without increasing the
differential order of the equations, which, on the contrary, would require additional boundary
conditions. It is not necessary to introduce a mapping function of the nonuniform grid in the
physical domain into a uniform grid in an infinite domain. \textit{A priori} tests on the turbulent channel
flow ($Re_\tau=180$ and $590$) highlight the approximation capability of the present procedure. A numerical
example is given, which draws attention to the nonlocal effects on the solution due to the lack of
noncommutation terms in the motion equation and to the efficiency of the present procedure in
reducing the commutation error on the solution.

\end{abstract}

\maketitle

\section{Introduction\label{par.introduction}}

The problem of the non commutativity of the filtering operation 
has been considered by Ghosal and Moin (1995)\cite{gm95}, van der Ven (1995)\cite{vdv95}, Fureby and Tabor (1997)\cite{ft97}, Vasilyev \textit{et al.} (1998)\cite{vlm98} and reviewed by Ghosal (1999)\cite{ghosal99} and Sagaut in his monography on large eddy simulation (LES) for incompressible flows (2001)\cite{s01}.
 
Ghosal and Moin\cite{gm95}  showed that the commutation error is of the second order in the filter width. Introducing a filter definition built on the mapping function of the non uniform grid, they proposed a procedure that can be used whenever numerical schemes based on pseudo-spectral methods or on finite differencing of an order higher than the second order are employed. 
The non commutation term are expanded in a Taylor series in the filter width, where the coefficients depend on the spatial derivatives of the filtered field and the mapping function. In this way, extra terms appear in the filtered equations, which increase the differential order of the equations. The authors suggested
both the use of additional boundary conditions to mantain the well-posedeness of the problem or the use of asymptotic expansions of the filtered variables, in terms of the square of the filter width, which however requires the solution of additional non homogeneous perturbative problems.
 
A family of one parameter filters commuting with differentiation up to any given order in the filter width, which is assumed nonuniform in the integration domain, was constructed by van der Ven \cite{vdv95}. If a discretization scheme of a given order is adopted in a LES, one may select a filter inside the family so that the lack of commutation between differentiation and filtering can be neglected.

A general formulation of the commutation error, due to a non uniform filter
and to the presence of boundaries, has been proposed by Fureby and Tabor\cite{ft97}. Boundary domain terms are explicitly formulated inside this representation. A detailed numerical analysis of the field distribution of the intensity of the non commutation terms 
is given in this paper by comparing LES and direct numerical simulation (DNS) data obtained from simulations of the incompressible turbulent channel flow at $Re_\tau= 180$ and $395$. It has been found that the local intensity can be as high as $21\%$ of the local advection, with a field volume averaged relative intensity of about  $8\%$. Lower values apply if reference is made to the sum of the local advection, pressure gradient, and molecular viscous flux terms. An interesting result is that the use of different subgrid scale models negligibly affects the local and average values of the sum of the commutation error terms. As expected, the high relative intensities of the commutation terms are concentrated in the flow regions where the gradient of $\delta$ is high. However, the question of the possible extension of the effects on the macroscopic scale of the flow is still left open.

A generalization of the procedures proposed by Ghosal and Moin and van der Ven is presented in the paper by Vasilyev \textit{et al.}\cite{vlm98}. A minimization of the commutation error is achieved by using a class of filters with $n-1$ vanishing moments, where $n$ is the order of the employed numerical discretization scheme. The authors also supply a group of rules to construct 
discrete filters that commutes with differentiation up to any given order
inside complex domains.

The method here proposed relies on an approximation of the specific non commutation term that corresponds to the different terms of the motion equations. A commutation approximation of the fourth order in the filter width can be obtained thanks to the introduction of successive levels of average. 
See  Sec. \ref{par.non-comm} for the basic formulation relevant to an isotropic grid stretching and the Appendix for the more general anisotropic case, which also specifies the formulation that is appropriate to  wall-bounded flows.

While performing large eddy simulations, the present approach can conveniently be used together with subgrid models based on analog multi-level filtering, e.g. models which apply the dynamic procedure, Germano \textit{et al.}\ (1991)\cite{g91}, Germano (1992)\cite{g92} or the Bardina mixed  model (1980)\cite{bfr80}. 
The filtering approach we use in this paper is that of the very fundamental volume average, first applied by Smagorinsky in 1963\cite{sma}. The volume average formulation is advantageous because it does not introduce an error associated to domain boundaries, thus avoiding the problem of the addition of further non commutation terms in the equations.

%



The variable scale filtered  Navier-Stokes equations, including the commutation terms approximation, are given in Sec. \ref{par.non-comm-A}. 


A priori tests on the turbulent channel flow data bases by Alfonsi \textit{et al.}\ (1998)\cite{passoni1} and Passoni \textit{et al.}\ (1999)\cite{passoni2}, $Re_\tau=180$, and by Moser \textit{et al.}\ (1999)\cite{mkm1999}, $Re_\tau=590$, are presented in Sec.\ \ref{par.test-A}.
An example of application of the  numerical procedure is presented in Sec.\ \ref{par.test-B}, which focuses attention on the fact that this systematic error is important throughout the entire flow and not only in the regions where the non homogeneous terms of the motion equation, which originate from the lack of commutation of the operations of differentiation and filtering, are different from zero. The capacity of the present procedure to reduce the relevant absolute and relative  errors is shown.

Before proceeding to the other sections, it is necessary to open a  digression on the terminology adopted in what follows. Since, among the  points being discussed in the paper, there are: the structure of the motion equation, once a variable scale filtering is used, and the role played by the terms which originate from the non commutation filter-differentiation, we have had to clearly distinguish the concept of commutation error on the flow solution from that of  non commutation term in the equations.
By {\em commutation error} we mean the error which affects the flow solution, when a variable scale filter is used, but the equations are used as if the operations of filtering and differentiation commute.
By {\em non commutation term} we mean any of the terms which originate
in the equation of motion when the filter scale is a function of the point. It is necessary to recall that in previous literature the latter was called commutation error, since the equation were always used as if they were commutative and the omissiom of the terms, which should make them complete when the filter length varies, introduced the error in the solution.

\section{Non commutation terms and their approximation\label{par.non-comm}}

The loss of the commutation between the spatial filtering and the differentiation operations is related to the use of a variable filter,
which in the more general configuration is anisotropic. In this case the filter width is the vector $\mdel(\xg) = (\delta_1(\xg), \delta_2(\xg), \delta_3(\xg))$.

For reading convenience and as the isotropic stretching configuration is conceptually non reductive, a scalar filter scale $\delta(\xg)$ is assumed in what follows. However, the general anisotropic configuration of stretching is dealt with in the Appendix, which also specifies a filtering formulation that is suitable for wall-bounded flows.

Let us suppose we have chosen a given class of integration volumes
 
$$
V_{\delta}= \left\{ \etag \in \dR^3: \parallel \etag  \parallel \langle  \delta \right\}
$$

\noindent and  an average operation for the variable $f(\xg)$:

\begin{equation}
\langle f\rangle_{\delta} =\frac{1}{V_\delta} \int_{V_\delta} f(\xg + \etag) d \etag =
\frac{1}{V_1} \int_{V_1} f(\xg + \delta \xig) d \xig, \;\;\; 
\label{def-media}
\end{equation}
where the transformation $\xig=\etag/\delta$ has been used and, as a consequence, $V_1 = V_\delta / \delta^3$.
Please note that, with this choice, the width of the averaging volumes is twice the filter scale.

A variable filter scale is introduced by allowing  $\delta$ to be a function of point, $\delta=\delta(x)$. 
In this case
\begin{eqnarray}
\frac{\partial}{\partial x_{i}} \langle f\rangle_{\delta} &=&
\frac{\partial}{\partial x_{i}} [ \frac{1}{V_1} \int_{V_{1}} f(\xg + \delta(\xg)  \xig) d \xig ] \nonumber\\
&=&
\langle \nabla f \cdot \mdel_i\rangle_{\delta}
+ \frac{\partial \delta}{\partial x_i} (\xg)
\langle  \nabla f \cdot \xig \rangle
\label{derfilt}
\end{eqnarray}
\noindent By virtue of the fact that
\begin{eqnarray}
\frac{\partial}{\partial \delta} \langle f\rangle_{\delta} &=&
\frac{\partial }{\partial \delta } \left[ \frac{1}{V_1} \int_{V_{1}} f(\xg + \delta(\xg) \xig) d \xig \right]\nonumber\\
&=& \frac{1}{V_1} \int_{V_{1}} \nabla f \cdot \xig d\xig,
\label{der1}
\end{eqnarray}

\noindent  recalling that $\langle \nabla f \cdot \mdel_i\rangle_{\delta}
= \langle  \displaystyle \frac{\partial f}{\partial x_i}\rangle_\delta$, it results that the filter of the derivative is a
differential operator acting on the filtered field:

\begin{equation}
\langle \frac{\partial f}{\partial x_i}\rangle_{\delta} =  \frac{\partial}{\partial x_i} \langle f\rangle_{\delta} -
\frac{\partial \delta}{\partial x_i} \frac{\partial}{\partial \delta} \langle f\rangle_{\delta}
\label{filder}
\end{equation}

\vspace{3mm}
\noindent The non commutation term $\mcalC_i'$, which is defined as
\begin{equation}
\mcalC_i'(\langle  f\rangle_{\delta}) = \langle  \frac {\partial f}{\partial x_i}\rangle_{\delta} - \frac{\partial}{\partial x_i} \langle  f \rangle_{\delta}, 
\label{Commuerdef}
\end{equation}

\noindent can be represented through  (\ref{filder}) by the product of the filter space derivative and the filter derivative of the filtered variable: 
\begin{equation}
\mcalC_i'(\langle  f\rangle_{\delta}) = 
- \frac{\partial \delta}{\partial x_i} \frac{\partial}{\partial \delta} \langle f\rangle_{\delta}
\label{Commuer}
\end{equation}

The here proposed method is based on 
an approximation of relation (\ref{Commuer}). 
The problem could be faced  adopting a
truncated series expansion
of $\langle f\rangle_\delta$ in terms of powers of $\delta$ (Ghosal and Moin, 1995)\cite{gm95}. However, this would increase
the order of the equations, and thus require additional boundary conditions. Here a numerical approximation of the $\delta$ first derivative is used in conjunction with  truncated $\delta$ expansions. Let us write the second order finite difference approximation 

\begin{equation}
\frac{\partial \langle f\rangle_{\delta}}{\partial\delta}=
\frac{1}{2h}\left(\langle f\rangle_{\delta+h}-\langle f\rangle_{\delta-h}\right)+O(h^2)
\label{app-der0}
\end{equation}
Choosing
$h=\delta$ 

\begin{equation}
\frac{\partial \langle f\rangle_{\delta}}{\partial\delta}=
\frac{1}{2\delta}\left(\langle f\rangle_{2\delta}-\langle f\rangle_{0}\right)+O(\delta^2)
\label{app-der}
\end{equation}

\noindent Now, the problem to face is that of the approximation of $\fm_0=f$ and $\langle f\rangle_{2\delta}$ in terms of relevant averaged quantities.
Using a Taylor expansion of the integrating function in (\ref{def-media}), we obtain the following expression for $\fm_\delta$, in terms of $f$, and the filter width:
\begin{eqnarray}
\fm_\delta(\xg)&=& f(\xg) + \frac{1}{2} a_{1,0,0}\nabla^2 f(\xg)\delta^2 + \nonumber \\
&&+ \frac{1}{4!} \left[ a_{2,0,0} (\partial_{1}^4 +\partial_2^4+\partial_3^4)f(\xg) \right. \nonumber\\
&&+ \left. 6 a_{1,1,0} (\partial_{1}^2\partial_2^2+\partial_{1}^2\partial_3^2+\partial_2^2\partial_3^2)f(\xg)\right]\delta^4 + O(\delta^6) \nonumber \\
&=& f(\xg) +  F_1[f]\delta^2 +  F_2[f]\delta^4 + O(\delta^6)
\label{s1}
\end{eqnarray}
where coefficients $a_{ijk}$ are defined as
\begin{equation}
a_{ijk}=\frac{1}{V_1}\int_{V_1} \xi_1^{2i} \xi_2^{2j}\xi_3^{2k} {\rm d}\xig.
\label{eq.coeff-a}
\end{equation}
and the operators $F_1, F_2$ as:
\begin{eqnarray}
F_1[\cdot] &=&  \frac{1}{2} a_{1,0,0}\nabla^2 \cdot\\
\label{OF1}
F_2[\cdot] &=&  \frac{1}{4!} \left[ a_{2,0,0} (\partial_1^4 +\partial_2^4+\partial_3^4) \cdot + 6 a_{1,1,0} (\partial_1^2\partial_2^2\right.\nonumber\\
&&+\left.\partial_1^2\partial_3^2+\partial_2^2\partial_3^2)\; \cdot \;\right] \label{OF2}
\end{eqnarray}
From (\ref{s1}) it follows that
\begin{equation}
f=\fm_{\delta}- F_1[f] \delta^2 + O(\delta^4)
\label{s2}
\end{equation}
and then, averaging (\ref{s2}) on a volume of linear dimension $2\delta$, 
\begin{equation}
\fm_{2\delta} = \langle \fm_{\delta}\rangle_{2\delta}-  \langle  F_1[f]\delta^2\rangle_{2\delta} + O(\delta^4).
\label{eq.duedelta}
\end{equation}

However, from (\ref{s1}) it can be observed that
\begin{eqnarray*}
\langle  F_1[f]\delta^2\rangle_{2\delta} &=& F_1[f]\delta^2 + 4 F_1 [F_1[f] \delta^2] \delta^2+ ... =\\
&=&\delta^2 F_1 [f] + O(\delta^4)
\end{eqnarray*}
so that
\begin{equation}
\fm_{2\delta}=\langle \fm_{\delta}\rangle_{2\delta }- F_1[f]\delta^2 + O(\delta^4)
\label{2d}
\end{equation}
\noindent When the expressions (\ref{s2}) for $f$ and (\ref{2d}) for $\fm_{2\delta}$ are introduced into (\ref{app-der}), one obtains
\begin{equation} 
\frac{\partial \langle f\rangle_{\delta}}{\partial\delta}=
\frac{1}{2\delta}
\left(\langle \langle f\rangle_{\delta}\rangle_{2\delta}-\langle f\rangle_{\delta}\right)+O(\delta^2)
\label{appder} 
\end{equation}

\noindent When using (\ref{appder}), the non commutation term $\mcalC'_i$ [see (\ref{Commuer})] can be approximated by
\begin{equation} 
\tilde{\mcalC}'_{i}(\fm_{\delta}) = 
- \frac{\partial \delta}{\partial x_i}
\frac{1}{2\delta}
\left(\langle \langle f\rangle_{\delta}\rangle_{2\delta}-\langle f\rangle_{\delta}\right) 
\label{appcomm}
\end{equation}

\noindent which implies
\begin{equation}
{\mcalC}'_{i}(\fm_{\delta}) = 
\tilde{\mcalC}'_{i}(\fm_{\delta}) 
+ \frac{\partial \delta}{\partial x_i}O\left(\delta^2\right).
\label{app1}
\end{equation}

In order to analize the approximation error $(\partial_i \delta) O(\delta^2)$ and give a true estimate of it, let us write: 
\begin{equation}
\delta(\xg)=\Delta \varphi(\xg),
\label{deltaphi}
\end{equation}
where $\Delta$ is a reference value of the filter width which is usually associated to the portion of the domain where conditions of near homogeneity of the flow hold. 
Function $ \varphi(\xg)$ is a positive non dimensional function which belongs to the interval $[\delta_{min}/\Delta,1]$, in the homogeneous region of the flow $\varphi(\xg)$ is constant and equal to 1. Function $\varphi(x)$  varies in the inhomogeneous regions, though it keeps values that are greater than $\delta_{min}/\Delta$, which is a value that must correspond: (i) -- when the local scale invariance may be supposed -- to a convenient minimum value of the filter width still inside the inertial range, and (ii) -- when the local scale invariance does not hold -- to a scale of the order of the scale which characterises the local turbulence structure, as, in case of wall flows, is the scale of the quasi-streamwise vortices peculiar to the viscous sublayer (see  Moin and Kim, 1982\cite{mk82}, and Ghosal, 1999\cite{ghosal99}).

Introducing (\ref{deltaphi}) and $\partial_\delta=\Delta^{-1}\partial_\varphi$ into (\ref{Commuer}) and deducing the $\delta$ derivative from expansion (\ref{s1}), written up to the fourth order of accuracy, the non commutation term can be estimated 
\begin{eqnarray}
\mcalC'_{i}(\fm_\delta) &=& - \frac{\partial \varphi}{\partial x_i} \frac{\partial}{\partial \varphi} \fm_{\Delta\varphi} = \nonumber \\
&=& - 2\frac{\partial\varphi}{\partial x_i}
\varphi(\xg) \left(F_1[f] \Delta^2 \right.\nonumber\\
&&\left.+ 2 \varphi^2(\xg) F_2[f] \Delta^4 + O(\Delta^6) \right)
\label{commerrest}  
\end{eqnarray}
Using  expantion (\ref{s1}) twice, after having inserted (\ref{deltaphi}),  approximation (\ref{appcomm}) can be estimated as
\begin{eqnarray} 
\tilde{\mcalC}'_{i}(\fm_{\delta}) &=& - 2
\frac{\partial\varphi}{\partial x_i}
\varphi(\xg)\left(F_1[f] \Delta^2 + (4 \varphi^2(\xg)F_2[f]\right.\nonumber\\
&&\left.+ F_1[\varphi^2 F_1[f]])\Delta^4+O(\Delta^6) \right)
\label{appcomm1}
\end{eqnarray}
\noindent  To keep the validity of estimates (\ref{commerrest}) and (\ref{appcomm1}),
care must be taken to select a function $\varphi(\xg)$ which, in the region of filter variation, also   posseses first and second derivatives of $O(1)$.
Possible examples are trascendental functions such as $\arctan (\xg), \tanh (\xg)$.

\noindent A comparison of (\ref{commerrest}) and  (\ref{appcomm1}) yields
\begin{equation}
\mcalC'_{i}(\fm_\delta)-\tilde{\mcalC}'_{i}(\fm_\delta)\approx O(\Delta^4)
\label{fine1}.
\end{equation}

Consequently, when (\ref{appcomm}) is used, a fourth order non commutation term in (\ref{filder}) is produced instead of the second order error, which would be obtained by totally neglecting the lack of commutation \cite{gm95}. Introducing a finite difference approximation of a higher order than (\ref{app-der0}), 
and, consequently, further levels of average, it could be possible to increase the accuracy of the approximation of the non commutation term, leading to a higher order error in (\ref{fine1}).




This analysis pertains to differential operators of the first order. The analysis is similar for the second order differential operators. The structure of the correction terms remains the same and to reach the fourth order of accuracy the same number of levels of average must be mantained. The approximation of the non commutation term now includes the filter of the variable spatial first derivatives.

The non commutation term of the second derivatives, being defined by
\begin{equation}
\mcalC''_{ii}(\fm_\delta)= \langle \frac{\partial_2 f}{\partial x_i^2}\rangle - \frac{\partial^2}{\partial x_i^2} \fm_\delta,
\label{defin2}
\end{equation}
can be obtained by taking the derivative of the first derivative (\ref{derfilt}) as
%
\begin{eqnarray}
\mcalC''_{ii}(\fm_\delta) &=& 
-\frac{\partial^2\delta}{\partial x_i^2}\frac{1}{V_1}\int_{V_1}\sum_{j=1}^{3} \xi_j\frac{\partial f}{\partial x_j}(\xg+\delta\xig) {\rm d}\xig \nonumber\\
&&-2 \frac{\partial\delta}{\partial x_i} \frac{1}{V_1}\int_{V_1}\sum_{j=1}^{3} \xi_j\frac{\partial^2 f}{\partial x_j\partial x_i}(\xg+\delta\xig) {\rm d}\xig \nonumber\\
&&-\left(\frac{\partial\delta}{\partial x_i}\right)^2
\frac{1}{V_1}\int_{V_1}\sum_{j,k=1}^{3} \xi_j\xi_k \frac{\partial^2 f}{\partial x_j\partial x_k}(\xg+\delta\xig) {\rm d}\xig\nonumber\\
&&
\end{eqnarray}
that is,
\begin{eqnarray}
\mcalC''_{ii}(\fm_\delta )&=& - \frac{\partial^2\delta}{\partial x_i^2} \frac{\partial}{\partial \delta}\fm_\delta
-2 \frac{\partial\delta}{\partial x_i}\left(\frac{\partial^2}{\partial\delta \partial x_i}\fm_\delta\right) \nonumber\\
&& -
\left(\frac{\partial\delta}{\partial x_i}\right)^2 \frac{\partial^2}{\partial \delta^2}\fm_\delta
\label{commu2}
\end{eqnarray}
\noindent and can consequently be approximated using the finite difference for the $\delta$-derivatives, as performed for the non commutation term of the first derivatives. The use of the standard three-point formula for the second derivative of $\fm_\delta$ with respect to $\delta$ and of relation (\ref{appder}) for  
$\partial^2_{\delta x_i} \fm_\delta$ [second term on the right hand side of equation (\ref{commu2})],
yields
\begin{eqnarray}
\tilde{\mcalC}''_{ii}(\fm_{\delta}) &=& -
\frac{\partial_i\delta(x)}{\delta(x)} \left[ \langle \partial_i\fm_{\delta}\rangle_{2\delta}-\partial_i\fm_{\delta} \right] \nonumber\\
&&-
\frac{(\partial_i\delta(x))^2+\delta(x)\partial^2_{i}\delta(x)}{2\delta^2(x)} \left[ \langle \fm_{\delta}\rangle_{2\delta}-\fm_{\delta} \right],\nonumber\\
&&
\label{appr2}
\end{eqnarray}
\noindent whose order of accuracy can be determined by again using the trasformation $\delta(\xg)=\varphi(\xg) \Delta$.
If the expansions of the non commutation terms $\mcalC''$ is compared with its approximantion $\tilde{\mcalC}''$, it is seen that, also in this case, the error is $O(\Delta^4)$.

Even if the present analysis is based on the use of the volume averages, it can be observed that it remains valid in the case where a more general kind of filtering is adopted. A weight function $g(\xig)$, introduced in (\ref{def-media}),  only modifies the coefficients $a_{ijk}$ [Eq. (\ref{eq.coeff-a})], which should now be defined as
$$
a_{ijk}=\int g(\xig)\xi_1^{2i} \xi_2^{2j}\xi_3^{2k} {\rm d}\xig,
$$
\noindent However, the non commutation term (6) and its approximation remain unchanged, provided the weight function has a compact support. It is always possible to choose $\delta(\xg)$, so that the actual integration domain of the filter lies inside the flow domain (for instance setting a value that is lower than the distance from the wall of the first layer of grid points for the minimum  of $\delta(\xg)$). In this way, the compactness of the support prevents the error linked to the presence of  finite boundaries\cite{ft97}.
It should be noted that this procedure operates in the physical space and does not rely on the use of a mapping function of the non uniform grid. Centered volumes of average have been adopted [see Eq.\ (\ref{def-media})], even though they are not strictly necessary as far as the average process is considered. However, this choice shows two advantages: (i) physically, when the flow is incompressible,  the center of the volume of average is also the center of gravity and thus the point of application of the average momentum, (ii) analytically, it allows for a compact, and second order in $\Delta$, representation of the non commutations terms, which in turn are approximated by the present procedure with an accuracy of the fourth order. The choice of non centered volume of averages, which, in principle, is mathematically feasible,
yields a first order in $\Delta$ representation of the non commutation terms, which would also lead to a much more cumbersome analytical structure\cite{gm95,s01}. 


\subsection{Non commutation terms in the averaged incompressible Navier-Stokes equations\label{par.non-comm-A}}

Let us consider the incompressible Navier-Stokes equations written in the form:
\begin{eqnarray}
\partial_i u_i = 0\label{nscont}\\
\partial_t u_i + \partial_j (u_i u_j) + \partial_i p - \nu \,\partial^2_{jj} u_i = 0
\label{ns}
\end{eqnarray}

\noindent If a filter operator is applied the system  becomes
\begin{eqnarray}
\langle \partial_i u_i\rangle_{\delta} &=& 0
\label{cles}
\\
\langle \partial_t u_i\rangle_{\delta} + \langle \partial_j (u_i u_j)\rangle_{\delta} + \langle \partial_i p\rangle_{\delta} - \nu \,\langle \partial^2_{jj} u_i\rangle_{\delta} &=& 0.
\label{mles}
\end{eqnarray}

\noindent When $\delta$, the linear scale of filtering (see for instance the definition proposed in Sec.\ \ref{par.non-comm}) 
is not uniform in the flow domain, the averaging and differentiation operations no longer commute.
By introducing the subgrid turbulent stresses $R_{ij}^{(\delta)}=\ui_{\delta}\uj_{\delta} - \langle  u_i u_j \rangle_{\delta} $  and the non commutation terms ${\mcalC_i}', {\mcalC_{ii}}''$, for the first and second derivatives, as discussed in Sec.\ \ref{par.non-comm} for the isotropic filter configuration  and in the Appendix for the general anisotropic and the wall-bounded flow  configuration [see the isotropic relations (\ref{Commuerdef}), (\ref{Commuer}), (\ref{defin2}) and (\ref{commu2}); the anisotropic relations (\ref{anis-Commuerdef})--(\ref{anis-Commuer}), (\ref{anis-defin2}), (\ref{anis-appcomma}), (\ref{anis-appr2}); the wall anisotropy relations (\ref{anis-wall-bounded}), (\ref{anis-wall-bounded2})),
the averaged equations are written as
%
\begin{eqnarray}
\partial_i \ui_{\delta} = - \mcalC'_i(\ui_{\delta}) \;\;\;\;\;\;\;\; \;\;\;\;\;\;\;\; \;\;\;\;\;\;\;\; \;\;\;\;\;\;\;\;&& 
\label{clesc}\\
\partial_t \ui_{\delta} + \partial_j (\ui_{\delta}\uj_{\delta}) + \partial_i \p_{\delta} - \nu \,\partial^2_{jj} \ui_{\delta}  - \partial_j R_{ij}^{(\delta)}
=\nonumber \\ =
- \mcalC'_j (\langle  u_i \rangle_\delta\langle u_j \rangle_{\delta}) - \mcalC'_i (\p_{\delta}) + \nu\,\mcalC''_{jj} (\ui_{\delta})+ \mcalC'_j(R_{ij}^{(\delta)} ).\nonumber\\
&&
\label{mlesc}
\end{eqnarray}
\noindent
Together with what has been explainded in detail in Sec.\ \ref{par.non-comm} (see (\ref{appcomm}) and (\ref{appr2})) the present procedure approximates the non commutation term terms on the right hand side of (\ref{clesc}, \ref{mlesc}) with an accuracy of the fourth order. 
The correction terms can thus be represented by the following group of relations which are determined from the field information obtained through two successive average levels (the second being computed over a linear scale $2 \delta$):

\begin{equation}
\tilde{\mcalC}'_i(\ui_{\delta}) = - \frac{\partial_i\delta(x)}{2\delta(x)}\left[\langle \ui_{\delta}\rangle_{2\delta}-\ui_{\delta} \right]
\label{Commuerui}
\end{equation}

\begin{equation}
\tilde{\mcalC}'_j(\ui_{\delta}\uj_{\delta}) = - \frac{\partial_j\delta(x)}{2\delta(x)}\left[\langle \ui_{\delta}\uj_{\delta}\rangle_{2\delta}-\ui_{\delta}\uj_{\delta} \right]
\label{Commuercon}
\end{equation}

\begin{equation}
\tilde{\mcalC}'_i(\p_{\delta}) = - \frac{\partial_i\delta(x)}{2\delta(x)}\left[\langle \p_{\delta}\rangle_{2\delta}-\p_{\delta} \right]
\label{Commuerp}
\end{equation}

\begin{eqnarray}
\tilde{\mcalC}''_{jj}(\ui_{\delta}) = - \sum_{j=1}^{3}\left\{
\frac{\partial_j\delta(x)}{2\delta(x)} \left[ \langle \partial_j\ui_{\delta}\rangle_{2\delta}-\partial_j\ui_{\delta} \right] +
\right. \nonumber \\ \left.
+\frac{(\partial_j\delta(x))^2+\delta(x)\partial^2_{j}\delta(x)}{2\delta^2(x)} \left[ \langle \ui_{\delta}\rangle_{2\delta}-\ui_{\delta} \right] \right\}
\nonumber\\
&&
\label{Commuerdiff2}
\end{eqnarray}

\begin{equation}
\tilde{\mcalC}'_j (R_{ij}^{(\delta)}) = - \frac{\partial_j\delta(x)}{2\delta(x)}\left[\langle R_{ij}^{(\delta)}\rangle_{2\delta}-R_{ij}^{(\delta)} \right], \label{CommuerSubgridstress}
\end{equation}
\noindent which must be accordingly modified in the case of anisotropy of the stretching of the computational grid, see the previous comments and the Appendix [(\ref{anis-appcomma})--(\ref{anis-appcommb}), (\ref{anis-appcommb2})--(\ref{anis-wall-bounded}) and (\ref{anis-appr2})--(\ref{anis-wall-bounded2})].

The adoption of the volume average allows the filtered variables to be fully supported inside the physical domain. As a consequence, a peculiar property of the present procedure is that there is an absence of non commutation terms associated to a finite or semi-infinite computational domain in the filtered equations. Such terms arise when the filtering operator requires the extension of the dependent variables beyond the rim of the domain [see Fureby and Tabor (1997)\cite{ft97}]. 

It can be observed that the  use of the double level of average highlights the convenience of coupling this procedure to subgrid models which also employ it. These subgrid models are the mixed model, by Bardina \textit{et al.}\ (1980)\cite{bfr80} and, in general, all models that apply the dynamical procedure, by Germano \textit{et al.}\ (1991)\cite{g91} and Germano (1992)\cite{g92}.

The filtered equations (\ref{cles}, \ref{mles}) are invariant under Galileian transformations. Under transformation $t' \rightarrow t, \, x'  \rightarrow x + ct$,  a spatial variation of the filter scale $\delta(x)$, in the $x,t$ reference system, becomes a spatio-temporal variation $\delta(x) = \delta(x'-ct') = \delta(x',t')$ in the $x',t'$ system.
The temporal dependence of the filter scale yields  to the presence of a non commutation term which is also associated to the non stationary term.  The trasformed non commutation terms, and the relevant approximations, released by the unsteady and the convective terms, however cancel each other. Furthermore, the terms obtained from the divergence of the stress tensor in a system like (\ref{clesc}, \ref{mlesc}) and the  corresponding approximations (\ref{Commuerp}) - (\ref{CommuerSubgridstress}) are all Galileian invariants, which assures that the variable scale filtered equations and their approximations are also such. 

Another general implication, linked to the presence of a finite domain, is that the boundary conditions for the filtered variables should be different from those for the unfiltered variables.
The problem of wall boundary conditions for the filtered field could be treated with this procedure by adopting
one of the classical approximated conditions, which rely on the introduction of a special subgrid model, that is, the wall model, which is apt to represent the inner layer dynamics and which puts the first grid point inside the logarithmic layer\cite{d70}$^,$\cite{s75}$^,$\cite{g87}$^,$\cite{pfm89}.
It could also be  treated  by  placing  grid points 
well inside the viscous sublayer to resolve the near-wall dynamics and  by assuming no slip and impermeability boundary conditions.
It should be recalled that the latter conditions, which in theory should not be used for filtered velocities,
introduce an error of $O(\Delta^2)$, independently of the filter shape. 
Their use requires the subgrid model, which should represent the non homogeneous and anisotropic structure of the viscous and buffer layers, to be altered in the inner region. 
The employement of anisotropic models based on a tensorial turbulent viscosity would be opportune, see Horiuti (1990)\cite{h90}, Carati and Cabot (1996)\cite{cc96}, the review monography by Sagaut (2001, Chap.\ 5.3)\cite{s01} and also the differential angular momentum model (Iovieno and Tordella, 2002)\cite{it02}, which, being based on the representation of the turbulent viscosity through the moment of momentum vector, is well suited to assume an anisotropic formulation.

\section{Numerical tests\label{par.test}}

\subsection{A priori tests on the turbulent channel flow\label{par.test-A}}

In this section a set of a priori tests  is presented, which provides information on the field distribution of the non commutation terms, their relevant approximations and their ratios with respect to the physical terms from which they arise. The data correlating the approximated and exact non commutation terms have been determined by filtering  the direct numerical simulation of the turbulent plane channel flow at $Re_{\tau}=180$, as performed by Alfonsi \textit{et al.}\ (1998)\cite{passoni1} and Passoni \textit{et al.}\ (1999)\cite{passoni2}, and at $Re_\tau=590$, as performed by Moser \textit{et al.} (1999)\cite{mkm1999}.  The longitudinal momentum balance, which implies a zero pressure non commutation term, is considered. Repetition on two different grid levels has been performed.
All the data here presented have been averaged over an interval of 1.2 revolution times.

\begin{figure}
\vspace{-11mm}
\includegraphics[width=\columnwidth]{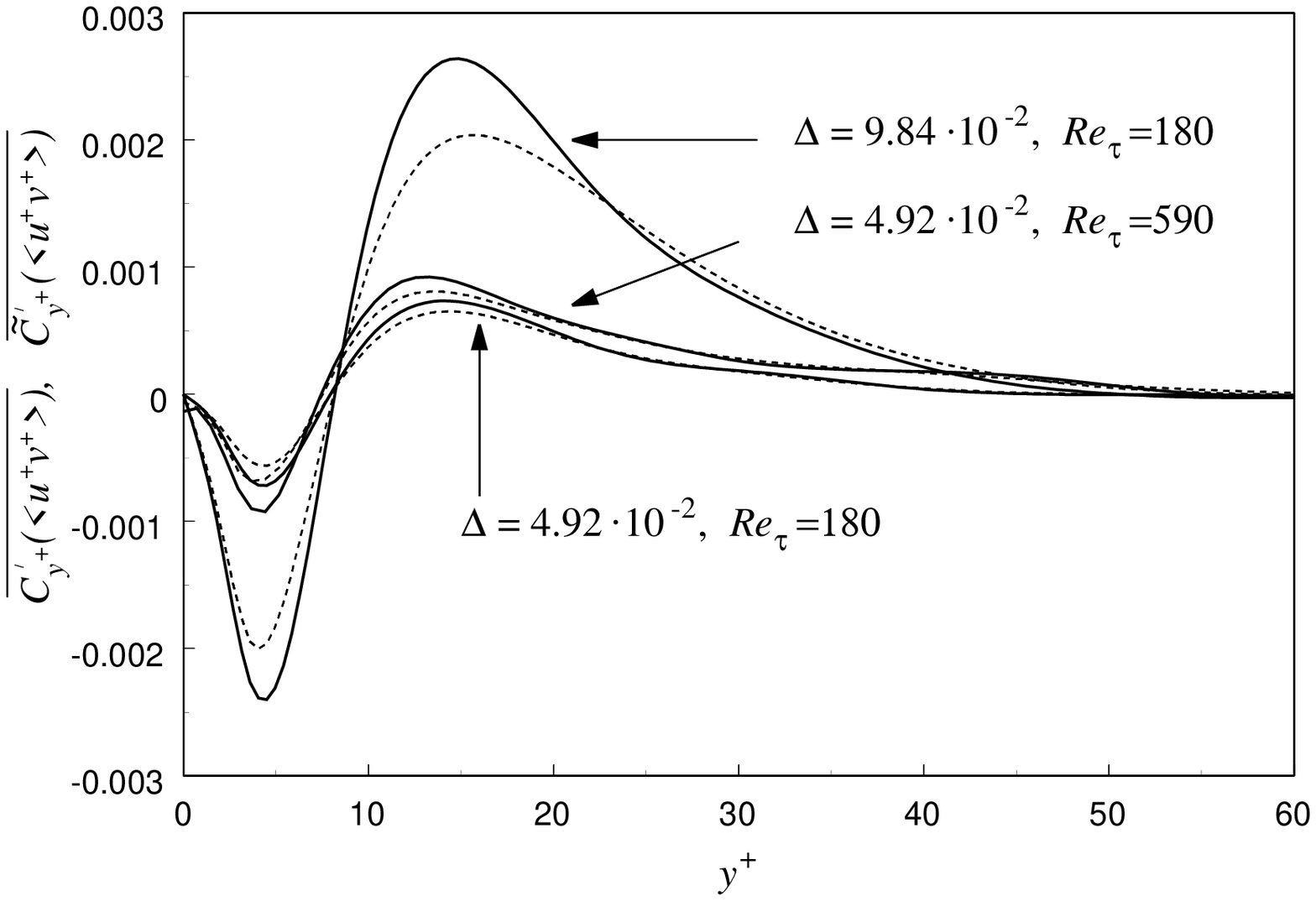}\\[-10mm]
\includegraphics[width=\columnwidth]{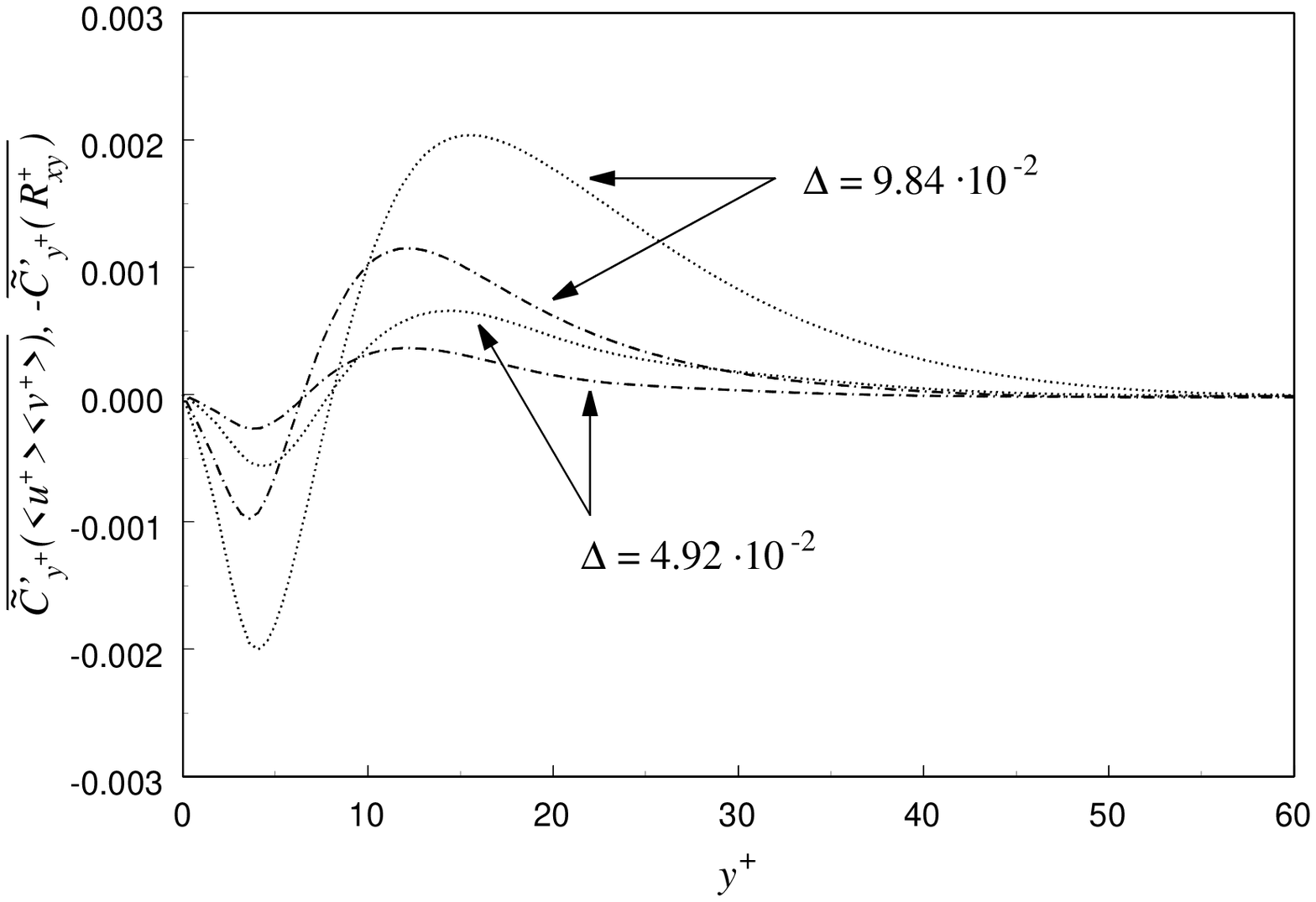}\\[-10mm]
\includegraphics[width=\columnwidth]{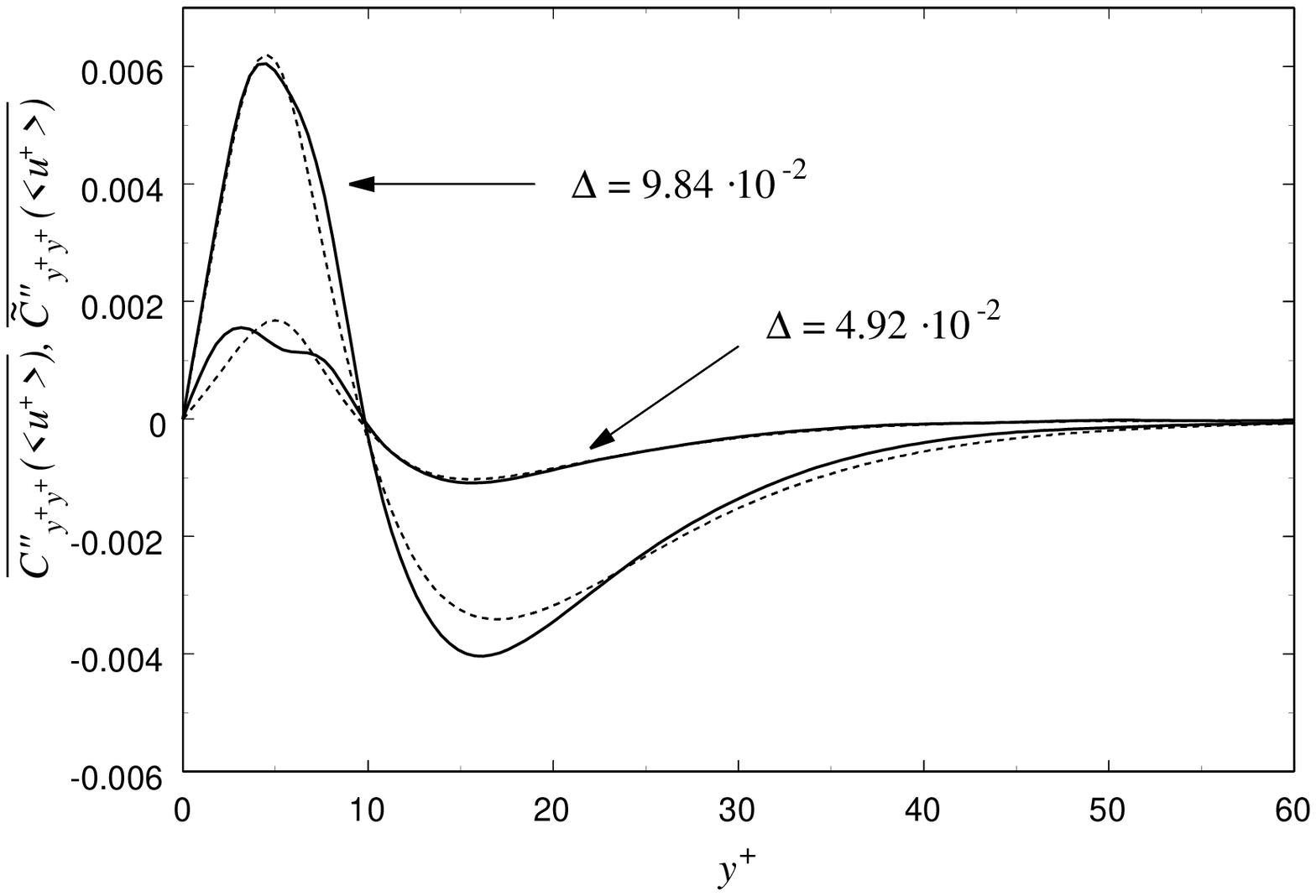}
\vspace*{-8mm}
\caption{(a): ({\bf -----}) Average values of the exact non commutation convection term and ({\bf -- -- --})  correspondent average values predicted by the procedure.
(b): Noncommutation approximated terms  for the resolved Reynolds stresses ({\bf -- . -- . --})
and for the subgrid scale stresses ({\bf --  --  --}). 
(c): ({\bf -----}) Average values of the noncommutation diffusive term and ({\bf -- -- --})  correspondent values predicted by the procedure.} 
\label{fig.1}
\end{figure}

Figure \ref{fig.1} shows the distributions of the exact non commutation terms and of their approximation according to the present procedure,  at two filtering
levels.  The convection term of the longitudinal momentum balance, its grid and subgrid scale decomposition and the diffusive term  are shown in parts (a), (b) and (c), respectively.
The filter $\mdel (\xg) = (\Delta x, \varphi(y) \Delta y, \Delta z)$, with constants $\Delta x, \Delta y$ and $ \Delta z$, varies along the transversal  non dimensional $y$ direction according to $ \varphi(y)  $, where $\varphi(y) \in [0, 1]$, and $y \in [-1 , 1]$
is a function of at least class $\textbf{C}^2$. The variation of $\varphi(y)$ has been laterally arranged (in 20\% of the channel width along the walls) as follows
\begin{equation}
\varphi(y) = \frac{\tanh a(y+1)\tanh a(1-y)}{\tanh^2 a},\;  a=4,
\label{phi1}
\end{equation}
\noindent where $a$ is the parameter that controls the gradient of the filter scale at the wall. The non commutation terms on the first and second derivatives have been determined for such an anisotropic structure of the filter through the use of relations  (\ref{anis-wall-bounded}) and (\ref{anis-wall-bounded2}). The data in Fig.\ 1 (a, c) show that the present procedure yields, on average,  for the Reynolds stress
$$ \frac{\mid\mcalC'\mid_\Delta}{\mid\mcalC'\mid_{2\Delta}}=   0.26\;,
\frac{\mid\mcalC'-\tilde\mcalC'\mid_\Delta}{\mid\mcalC'-\tilde\mcalC'\mid_{2\Delta}}= 0.13\;,
$$
$$
\frac{\mid\mcalC'-\tilde\mcalC'\mid_\Delta}{\mid\mcalC'\mid_{\Delta}}=  0.09 ,
\frac{\mid\mcalC'-\tilde\mcalC'\mid_{2\Delta}}{\mid\mcalC'\mid_{2\Delta}}=  0.18 ,$$
and for the viscous stress
$$ \frac{\mid\mcalC''\mid_\Delta}{\mid\mcalC''\mid_{2\Delta}}=   0.25\;,
\frac{\mid\mcalC''-\tilde\mcalC''\mid_\Delta}{\mid\mcalC''-\tilde\mcalC''\mid_{2\Delta}}= 0.17  \;,
$$
$$
\frac{\mid\mcalC''-\tilde\mcalC''\mid_\Delta}{\mid\mcalC''\mid_{\Delta}}=  0.10 ,
\frac{\mid\mcalC''-\tilde\mcalC''\mid_{2\Delta}}{\mid\mcalC''\mid_{2\Delta}}=  0.15 ,$$
\noindent where $\Delta=4.92\cdot 10^{-2}$.


These figures have been obtained by neglecting the data that belong to the first 5\%\ near the wall, where the numerical uncertainty due to the spatial discretization is high - especially as regard the exact non commutation terms computations - and where these results deteriorate by nearly 25\%.

The results of the numerical test at $Re_\tau= 590$ as compared with those at $Re_\tau= 180$, see Fig.\ \ref{fig.1} (a), show a good invariance  of the procedure accuracy with respect to  variations of the flow control parameter.

It should be remarked, that a procedure capable to predict at worst, using the rather large value of  $\delta=4.92 \cdot 10^{-2}$, the 90\% of the value of the non commutation terms may be considered accurate. In fact, the analysis of numerical errors in LES of Turbulence with cutoff in the inertial range (errors due to spatial discretization: finite-differencing errors and aliasing errors) shows that {\it the resulting  errors are very large, of the same order  and even larger than the magnitude of the subgrid term over most of the wavenumber interval,  for finite difference schemes up to eighth-order accurate, irrespective of the grid resolution} (cf. Ghosal S., 1996, pp.\ 201-202)\cite{ghosal96}. In such a general situation, it may be considered it a success that the procedure is capable of predicting nine-tenth of the value of the non commutation terms. The relevant average error cannot spoil the overall numerical reliability of the simulations since it is about one order of magnitude lower than the errors due to the spatial discretation.

A warning is necessary regarding the numerical computation of the exact values of the non commutation terms. Direct computation through the definition is not recommended, as, even with the implementation of a numerical differentiation of the sixth order of accuracy, it artificially amplifies the fluctuations that are naturally present in the data field. The exact values of the non commutation terms  should be correctly evaluated by using the integral representation of the derivatives ${\partial}/{\partial \delta}$ or
${\partial}/{\partial \delta_j}$, such as  (\ref{der1}) in Sec.\ \ref{par.non-comm} or (\ref{anis-der1}) in the Appendix. No such numerical problems affect the computation of the approximated non commutation terms. 

Figure \ref{fig.2} provides information on the relative importance of the exact  non commutation terms with respect to the physical terms which causes them.
In part (a) the average ratio ${\displaystyle |{\mcalC}^{'}_y(\langle  u v \rangle) / \partial_y (\langle  u v \rangle)}|$ has been plotted for $\Delta=4.92\cdot 10^{-2}$, $\Delta=9.84\cdot 10^{-2}$ and also with an increase of the wall value of the stretching factor $a=\partial_y \varphi=8$. 
It can be seen,
that close to the point where the Reynolds stress reaches its maximum ($y^+ \approx 30$), and the  divergence therefore takes in the average very small values,  ratio values as high as $0.38$ are reached with the coarser grid.
The comparison of these results with the results of Fig.\ \ref{fig.1}(b) indicates that 
the average exact value of the convection non commutation term is of the same order than the average divergence of the subgrid stresses.   Furthermore, it is interesting to observe that the doubling of the stretching factor $a$ increases the relevance of the noncommutation terms in this region nearly as much as the doubling of the grid coarsening does, see in Fig.\ \ref{fig.2}(a) the near wall region where $y^+ \le 14$. Figure \ref{fig.2}(a) also shows a positive comparison of the field integral value of ${\displaystyle |{\mcalC}^{'}_y(\langle  u v \rangle) / \partial_j (\langle  u u_j \rangle)|}$ given by Fureby and Tabor\cite{ft97} with the distribution of the same ratio that has been yielded by the database  used here.\cite{passoni1}$^-$\cite{passoni2}

With respect to the average ratio ${\displaystyle |{\mcalC}^{''}_{y y} (\langle  u \rangle) / \partial^2_{y y} (\langle  u \rangle)|}$, Fig.\ \ref{fig.2}(b) yields  maxima local values of about 100\% for the coarser resolution, and about 60\% for the finer one, close to where the relevant non commutation terms reach their local maxima near to the wall. Leaving aside local maximum values detached from the wall and relevant to the coarser resolution, this ratio, in the central part of the field,  settles to constant lower values close to $0.2 \pm 0.1$.

\subsection{Commutation error on an analytical solution\label{par.test-B}}

\begin{figure}
\vspace{-5mm}
\includegraphics[width=\columnwidth]{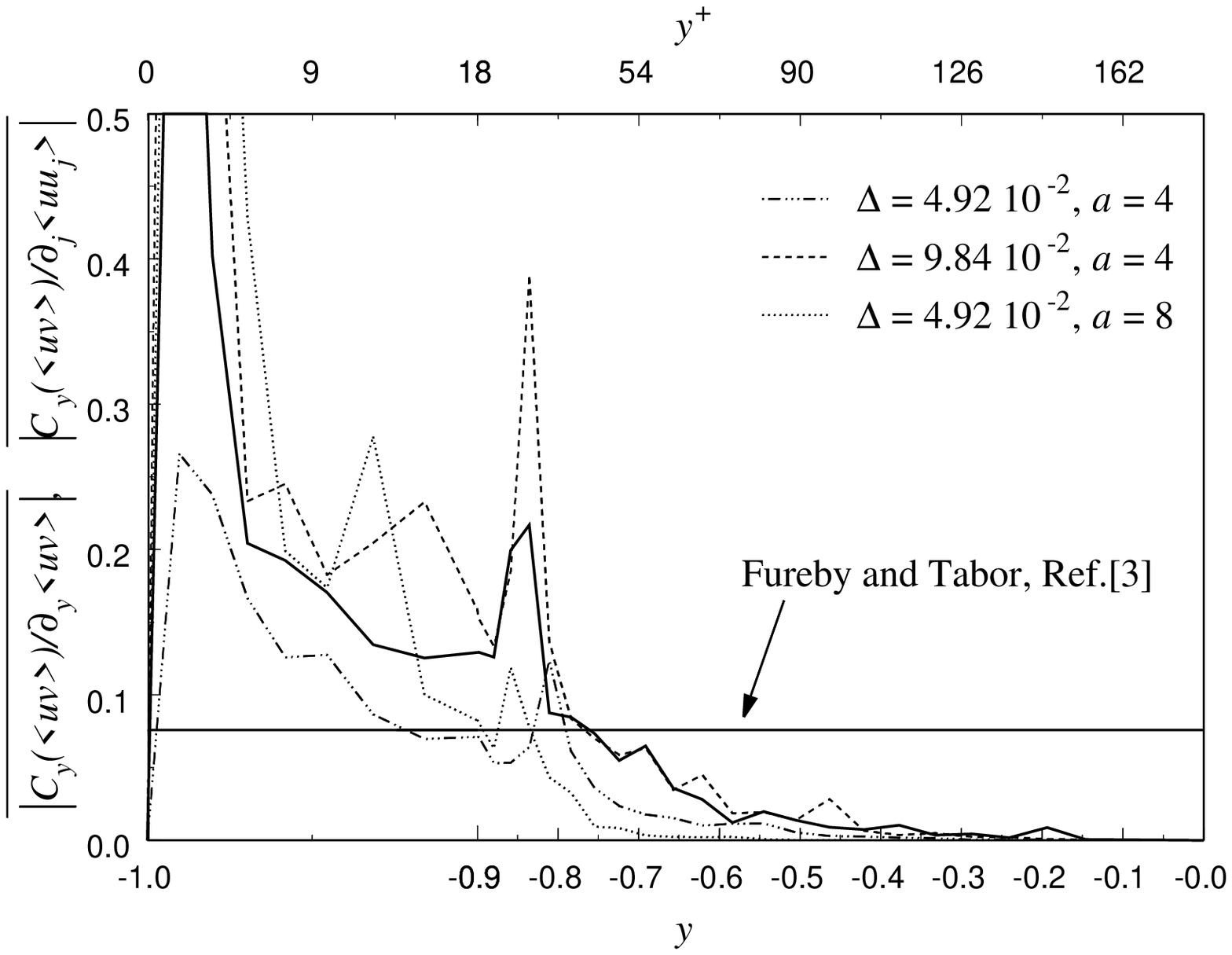}\\[-6mm]
\includegraphics[width=\columnwidth]{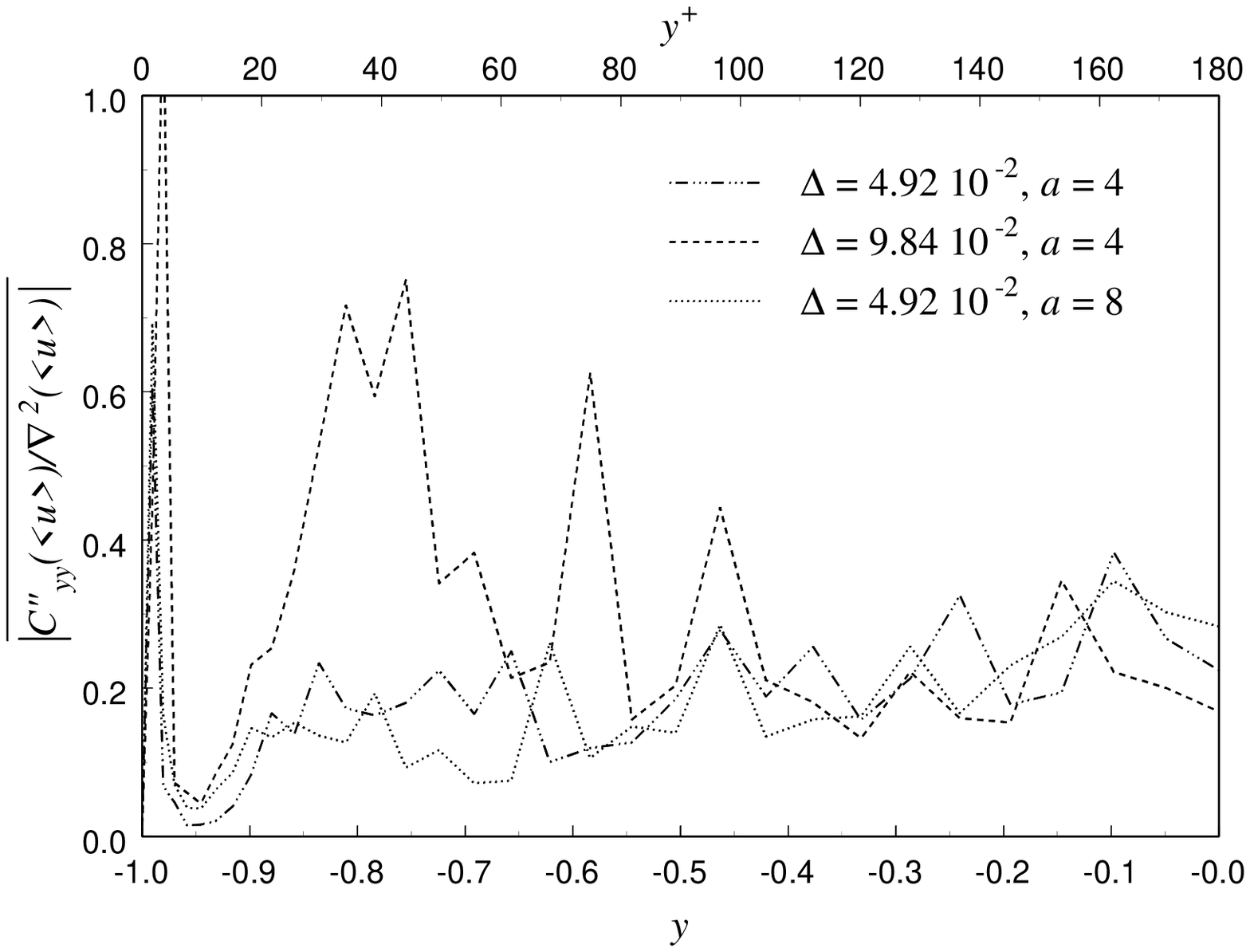}
\vspace*{-7mm}
\caption{Average of the absolute value of the ratio between the exact non commutation terms and the physical correspondent terms in the motion equation. (a) Convection, ({\bf -----}) ${\displaystyle |{\mcalC}^{'}_y(\langle  u v \rangle) / \partial_j (\langle  u u_j \rangle)|}$, (b): diffusion. Adimensionalization by means of channel semi-width and $u_{\tau}$.}
\label{fig.2}
\end{figure}

As seen in Sec.\ \ref{par.non-comm-A},
there are four type of non commutation terms in the  variable scale filtered incompressible Navier-Stokes equations (\ref{clesc}), (\ref{mlesc}), which are all source terms. If one limits the analysis of their influence on the flow solution
to the determination of the field distribution of the values they take with regards to the values taken by the original terms of the equations, one would mainly find the foreseen result that the non commutation terms are not negligible where the gradient of the filter scale is high. This is however not sufficient to understand the way, localized rather than extended, in which  the commutation error affects the flow solution.

For this purpose, it has here been considered useful to study the behaviour of an extremely simple flow model, a sort of conceptual model, which has two characteristics: -- just one type of commutation source term is present, -- its exact filtered solution, that is, the variable scale filtered solution not affected by the commutation error, is known and thus could be used as the reference solution. The second characteristics
can only be obtained by filtering the exact solution of the unfiltered equation of the motion. 

On the other hand, to prove the efficiency of any given procedure for the correction of the commutation error, 
it is also necessary to know the exact filtered solution of a test flow, which, in turn, requires the knowledge of the exact flow unfiltered solution. 

Such a reference state cannot be found in a turbulent configuration of flow, for which no exact solution is available. Reference is  therefore made to a laminar flow, which has an exact solution. One should note that, in such a case, the filtered equation of the motion, when the filter length is a function of the point but the  commutation error correction is not considered, is identical to the unfiltered equation. 

The  steady laminar incompressible channel flow
has been selected as the test flow, since it has only one commutation term, the diffusion one, see (\ref{Commuerdiff2}), and its solution is analytically known. 

The  non dimensional  momentum equation for the steady incompressible channel flow is written as
\begin{equation}
\partial^2_y  u =  Re \, \partial_x p =  - \frac{(G \rho)^{1/2} d^{3/2}}{\mu},
\;\;\; \forall \, y \in (0,1)
\end{equation}
%

%
%
\begin{figure}
\vspace{-7mm}
\includegraphics[width=\columnwidth]{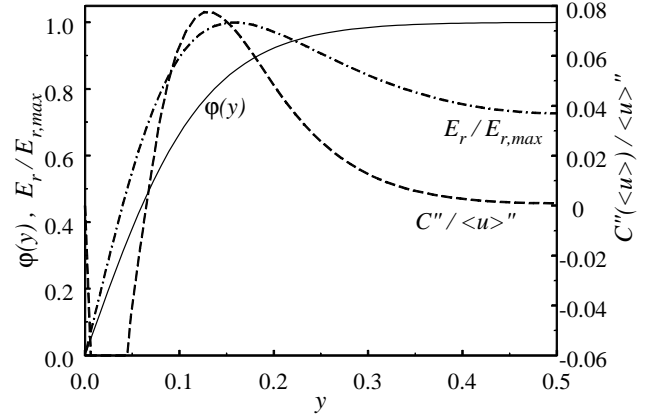}
\vspace{-7mm}
\caption{({\bf -----}) Filter scale across the channel ($\delta = \varphi(y) \Delta$, $a=8$, see (\ref{phi2}), III B);  ({\bf - - - -}) local values of the commutation term $\tilde{\mcalC}''$, see (\ref{Cfce}),  referring to the diffusion term; ({\bf----- - -----}) $R = E_r/E_{r\,max}$, relative commutation error for the solution of the filtered non corrected equation ($\mcalC''=0$) normalized with respect to the field peak value, $E_r = [(\uu - \uu_{exact})/\uu_{exact}]$.}
\label{fig.3}
\end{figure}
\begin{figure}[bth]
\vspace{-5mm}
\includegraphics[width=\columnwidth]{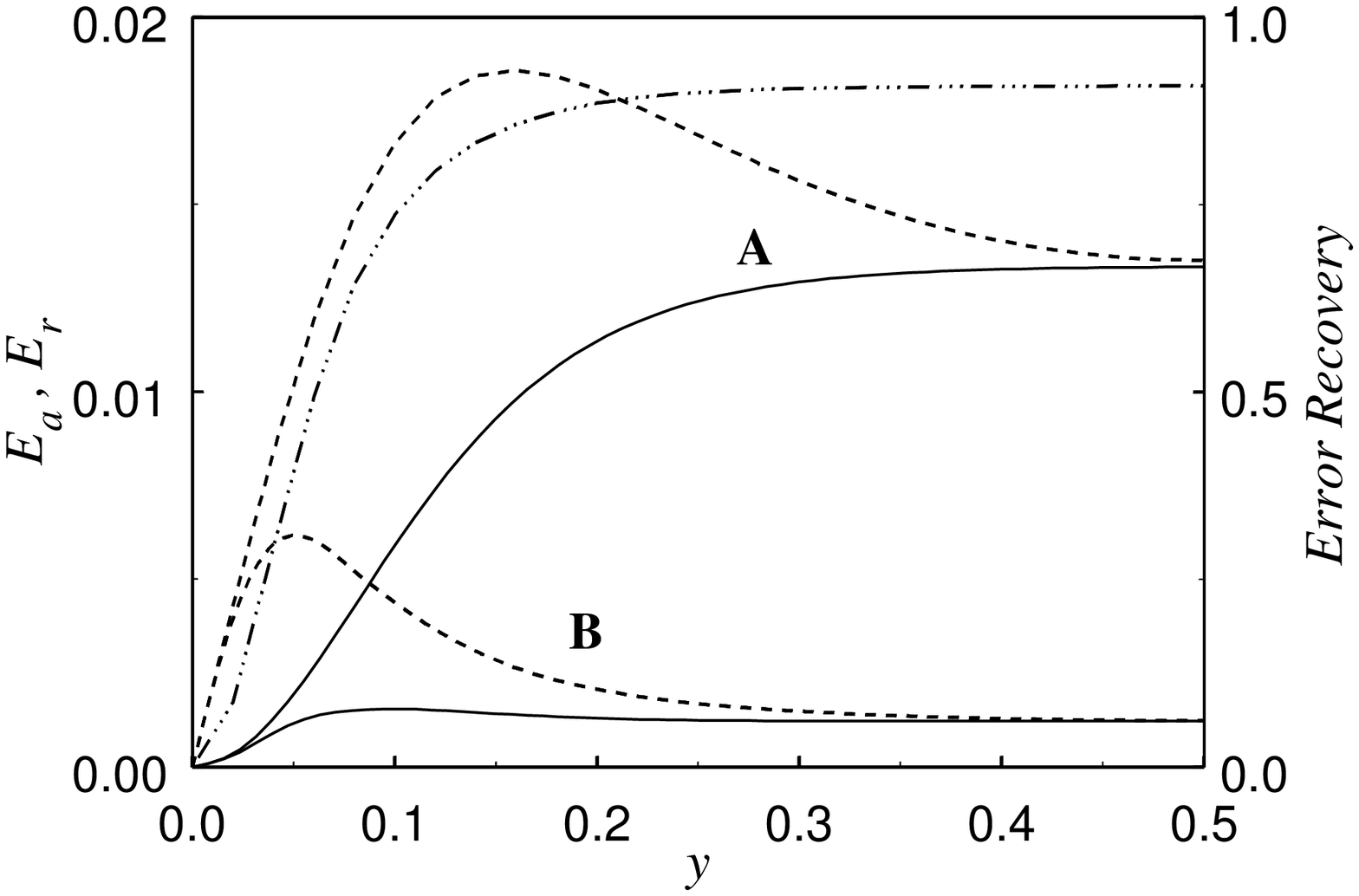}
\vspace*{-8mm}
\caption{Absolute $E_a = (8/Re)(\uu - \uu_{exact})$ ({\bf -----}) and relative $E_r = [(\uu - \uu_{exact})/\uu_{exact}]$ ({\bf - - - -}) error distributions of the filtered velocity distributions: {\bf A} -- without the commutation correction, {\bf B} -- with the commutation correction. Curve ({\bf----- - -----}): error recovery $(E_a({\bf A}) -E_a({\bf B}))/E_a({\bf A})$ with the distance from the wall.
The exact filtered velocity reference distribution is $\uu_{exact} = (Re/2)[y(1-y)- \frac{1}{3}\delta^2(y)]$.}
\label{fig.4}
\end{figure}
\noindent where $G$ is the modulus of the dimensional longitudinal pressure gradient
%
and the adimensionalization is based on the channel width $d$, the gradient $G$ and the density $\rho$ (the non dimensional pressure gradient $\partial_x p$ results equal to $-1$).
The boundary conditions are
\begin{equation}
u(0)=0, \;\;\; u(1)=0
\end{equation}
The corresponding filtered equation is
\begin{equation}
\partial^2_y \uu_{\delta}(y) + Re = -\mcalC''(\uu_{\delta}), \;\;\; \forall \, y \in (0,1)
\label{fce}
\end{equation}
\noindent where $\mcalC''(\langle u \rangle_{\delta})$ is set to zero to determine the solution which neglects the commutation error and where $\mcalC''(\langle  u \rangle_{\delta})$ is approximated by
\begin{eqnarray}
\tilde{\mcalC}''(\uu_{\delta})&=& - \frac{\delta'(y)}{\delta(y)} \left[ \langle \uu'_{\delta}\rangle_{2\delta}-\uu'_{\delta} \right]\nonumber\\
&& -
\frac{\delta'(y)^2+\delta(y)\delta''(y)}{2\delta^2(y)} \left[ \langle \uu_{\delta}\rangle_{2\delta}-\uu_{\delta} \right]
\label{Cfce}
\end{eqnarray}
\noindent to determine the solution which accounts for the commutation error with the present procedure.
The numerical solution of equation (\ref{fce}) is determined by solving the corresponding unsteady filtered equation                                                                                                                                                                                                                                                                                                                                                                                                    
\begin{equation}
\partial_t \uu_{\delta} - \frac{1}{Re} \partial^2 _y\uu_{\delta} = 1 + \frac{1}{Re} \tilde{\mcalC}''(\uu_{\delta}), \;\;\; \forall \, y \in (0,1)
\end{equation}
\noindent through a fourth order Runge-Kutta time integration scheme -- carried out until the steady state is reached -- coupled to a fourth order finite-difference discretization of the domain. The double level of average has
been computed using a third order Hermitian quadrature formula. 

The filter of the exact solution gives the velocity distribution ${\displaystyle \uu_{exact} = \frac{Re}{2}[y(1-y)- \frac{1}{3}\delta^2(y)]}$, which constitutes the reference on which the constrast between the commutation corrected filtered solution and the non corrected filtered solution is based. The filter scale varies along the transversal direction according to $\delta(y) = \varphi(y) \Delta $, where $\varphi(y) \in [0, 1]$, with $\Delta = 0.1$, is a function of at least class $\textbf{C}^2$. The variation of $\varphi(y)$ 
\begin{equation}
\varphi(y) = \frac{\tanh 2a y\tanh 2a(1-y)}{\tanh^2 a},
\label{phi2}
\end{equation}
\noindent has been laterally arranged in 20\% of the channel width along the wall, setting the parameter $a = 4$, see Fig.\ \ref{fig.3}. 

By contrasting the filtered exact solution,  the absolute and relative  errors relevant to the corrected and non corrected solutions are compared in Fig.\ \ref{fig.4}. The corresponding error recovery, with the wall distance, 
is also shown in Figs.\ \ref{fig.2} and \ref{fig.3}. It can be seen that the present procedure greatly reduces the commutation errors: in the central part of the flow, an almost full recovery  is obtained.  

For a comparison of the distributions of the local value of the non commutation term and of the relative
commutation error on the solution, see again Fig.\ \ref{fig.3}.
It is important to observe that, by neglecting the commutation correction, the field results to be affected by a systematic error not only in the region where the filter length varies, but also in the region where it is  constant.  This behaviour is due to the accumulation of errors on the velocity variable and its derivative, which is due to the lack of
the two diffusion addenda (see (\ref{fce}), (\ref{Cfce})) that should enter the momentum balance equation. Even through these terms are significantly different from zero in a limited portion of the flow they affect the entire field to a great extent. In the central part of the flow, where the non commutation term is very small (Fig.\ \ref{fig.3}), the relative commutation error results to be of the same order as the  peak value of the field, while the absolute error reaches its maximum value (see again Fig.\ \ref{fig.4}).

\section{Conclusions\label{par.concl}}

A procedure to explicitly insert the correction terms in order to counteract the commutation error associated to the use of a variable filter scale in the filtered equations of motion is here presented. With this procedure it is possible to directly compensate for the commutation error on the filtered field. The procedure uses  volume average filtering, but more general filter operators are also possible.  Both isotropic and  fully anisotropic filtering configurations are considered.
Approximated commutation terms, with an accuracy of the fourth order in the filter width, are inserted into the motion equations, which  do not increase their differential order. The difficulties related to the addition of further  boundary conditions are therefore avoided. The proposed representation of the commutator operators is based on truncated expansions in the filter width of finite difference approximations, that  make use of a multilevel average operation. This fact suggests the joint use 
of the present procedure with subgrid models which need an explicit filtering of the equations of motion, such as  dynamic and mixed models.

A set of a priori tests, with a plane channel flow DNS ($Re_{\tau}=180$) as a test field, proves the good correlation that the present procedure yields between the approximate and "exact" non commutation terms. It also provide  information on the relative importance 
of these terms with respect to the original physical terms. The influence of the field resolution on the general non commutation term is confirmed to be $O(\Delta^2)$. 
Asymptotically, the accuracy of the present approximation is expected to be  $O(\Delta^4)$. At the resolution levels corresponding to $Re_{\tau}=180$,
these tests show a reduction of the absolute errors, after halfing the reference filter scale, that is nearly $O(\Delta^3)$.

The filtering is a mathematical operation, that is not specific of the Navier-Stokes equations and is independent of the solution typology. When it is varied, the filter causes one kind of non commutation term for each differential term present in the equations of the motion. A test, for which the  analytically exact  (unfiltered and as a consequence filtered) solution is available, has been considered to overcome the limitation of an analysis, based  on the \textit{a posteriori}
determination of the relative order of magnitude of the non commutation terms, with regard to the original physical terms of the motion equations, and to  analyze the effects of the commutation error on a flow solution.
The chosen test flow is the two dimensional incompressible laminar channel flow, whose dynamics consist of the balance between the constant longitudinal pressure force and the viscous diffusion. In this case, only one type of commutation source term -- diffusion -- is present, and this is only of relative importance in the lateral part of the flow, according to the filter gradient dependence. The error, due to the lack of a non commutation term in the motion equation, is however also transferred to the central part of the flow.
The result is a biased filtered velocity distribution where the relative error in the central part of the flow, where the filter gradient is zero, is of the same order of magnitude as the local maximum error of the field, which is situated  at a distance from the wall of about $15\%$ of the channel width. 
It has been shown that the present  procedure can reduce the commutation error by  one order of magnitude in the central part of the field.


\begin{acknowledgments}
The authors would like  to thank Professor G.Passoni for helpful discussions and for making the turbulent channel flow database available. 
\end{acknowledgments}

\appendix*

\renewcommand{\appendixname}{APPENDIX}
\section{NON COMMUTATION APPROXIMATION FOR ANISOTROPIC FILTERS\label{appendice}}

When the geometry of the flow domain requires stretching  each direction independently, which implies $\mdel(\xg) = (\delta_1(\xg), \delta_2(\xg), \delta_3(\xg))$, it is opportune to adopt a class of integration volumes of the kind\\ 
\begin{equation}
V_{\mdel} = \left\{ \etag \in \dR^3: \parallel (\frac{\eta_1}{\delta_1}, \frac{\eta_2}{\delta_2}, \frac{\eta_3}{\delta_3} ) \parallel \langle  1 \right\}
\label{anis-vol-int}
\end{equation}

\noindent and  an average operation for the variable $f(\xg)= f(x_j + \delta_j \xi_j)$:\\
\begin{equation}
\langle f\rangle_{\mdel} =\frac{1}{V_{\mdel}} \int_{V_{\mdel}} f(\xg + \etag) d \etag =
\frac{1}{V_{\1}} \int_{V_{\1}} f(x_j + \delta_j \xi_j) d \xig, \;\;\; 
\label{def-media-anis}
\end{equation}
where $\1 = (1,1,1)$, the transformation $\eta_j=\delta_j \xi_j$ (with  det $\left({\partial \eta_i}/{\partial \xi_k}\right) = \delta_1 \delta_2 \delta_3$) has been introduced and thus $V_{\1} = V_\delta / \delta_1 \delta_2 \delta_3$.

In such a situation, by virtue of the fact that
\begin{eqnarray}
\frac{\partial}{\partial \delta_k} \langle f\rangle_{\mdel} &=& \frac{\partial }{\partial \delta_k} \left[ \frac{1}{V_{\1}} \int_{V_{\1}}  f (x_j + \delta_j (\xg) \xi_j) d \xig \right] \nonumber\\
&=& \sum_{j=1}^{3}  \frac{1}{V_{\1}} \int_{V_{\1}} \frac{\partial f}{\partial x_j}(x_j + \delta_j(\xg) \xi_j)
\times \nonumber\\
&&\times
\frac{\partial }{\partial \delta_k}(x_j + \delta_j (\xg) \xi_j) d \xig \nonumber\\
&=&  \frac{1}{V_{\1}} \int_{V_{\1}}  \xi_k \frac{\partial f}{\partial x_k} d\xig,
\label{anis-der1}
\end{eqnarray}
\noindent it is obtained
\begin{widetext}
\begin{eqnarray*}
\frac{\partial}{\partial x_{i}} \langle f\rangle_{\mdel} &=&
\frac{\partial}{\partial x_{i}} [ \frac{1}{V_{\1}} \int_{V_{\1}} f(x_j + \delta_j(\xg) \xi_j) d \xig ]
%
=
\sum_{j=1}^{3}  \frac{1}{V_{\1}} \int_{V_{\1}} \frac{\partial f}{\partial x_j}(x_j + \delta_j(\xg) \xi_j) \frac{\partial }{\partial x_i}(x_j + \delta_j (\xg) \xi_j) d \xig \\
&=&
\sum_{j=1}^{3}  \frac{1}{V_{\1}} \int_{V_{\1}} [\delta^{\rm K}_{ji} 
\frac{\partial f}{\partial x_j} + \frac{\partial \delta_j}{\partial x_i} \xi_j \frac{\partial f}{\partial x_j}] d \xig
%
=
\frac{1}{V_{\1}} \int_{V_{\1}}  \frac{\partial f}{\partial x_i} d \xig + 
\sum_{j=1}^{3} \frac{\partial \delta_j}{\partial x_i} \frac{1}{V_{\1}} \int_{V_{\1}}  \frac{\partial f}{\partial x_j} \xi_j d \xig
%
=
\langle  \frac{\partial f}{\partial x_i}\rangle_{\mdel}
+ \sum_{j=1}^{3} \frac{\partial \delta_j}{\partial x_i} \frac{
\partial \langle   f  \rangle_{\mdel}}{\partial \delta_j},  
\label{anis-derfilt}
\end{eqnarray*}
\end{widetext}
\noindent where $\delta^{\rm K}_{ji}$ is the Kronecker unit tensor.
\noindent As in Sec.\ \ref{par.non-comm} [see relation (\ref{filder})] for the isotropic configuration, it results that the anisotropic filter of the derivative is a differential operator that acts on the filtered field:
\begin{equation}
\langle \frac{\partial f}{\partial x_i}\rangle_{\mdel} =  \frac{\partial}{\partial x_i} \langle f\rangle_{\mdel} - \sum_{j=1}^{3}
\frac{\partial \delta_j}{\partial x_i} \frac{\partial}{\partial \delta_j} \langle f\rangle_{\mdel}
\label{anis-filder}
\end{equation}

\vspace{3mm}
\noindent The anisotropic non commutation term $\mcalC_i'$, which is defined as
\begin{equation}
\mcalC_i'(\langle  f\rangle_{\mdel}) = \langle  \frac {\partial f}{\partial x_i}\rangle_{\mdel} - \frac{\partial}{\partial x_i} \langle  f \rangle_{\mdel}, 
\label{anis-Commuerdef}
\end{equation}

\noindent can now be represented through  (\ref{anis-filder}) as the sum of the products of the filter space derivatives and the filter derivatives of the filtered variable: 
\begin{equation}
\mcalC_i'(\langle  f\rangle_{\mdel}) = 
- \sum_{j=1}^{3} \frac{\partial \delta_j}{\partial x_i} \frac{\partial}{\partial \delta_j} \langle f\rangle_{\mdel}
\label{anis-Commuer}
\end{equation}

Proceeding in strict analogy with what has been  done in Sec.\ \ref{par.non-comm}, the anisotropic non commutation term can be approximated through second order centered finite differences:
\begin{equation}
\tilde{\mcalC_i'}(\langle  f\rangle_{\mdel}) = - \sum_{j=1}^{3} \frac{\partial  \delta_j}{\partial x_i} \frac{1}{2 \delta_j}\left(\langle f\rangle_{\mdel + {\displaystyle \delta_j \eg_j}} - \langle f\rangle_{\mdel -{\displaystyle \delta_j \eg_j}} \right) + O(\delta_j^2)
\label{app-anis-der}
\end{equation}
\noindent Again, using a Taylor expansion of the integrating function in (\ref{def-media-anis}), where only even derivatives appear since the  domain of integration (\ref{anis-vol-int}) is symmetric with respect to all the integration variables,  expressions are obtained for quantities such as $\fm_{\mdel}$ and
$\fm_{\mdel \pm {\displaystyle \delta_j e_j}}$, in terms of $f$, and the filter width -- which is now defined as $\delta_j = \Delta \varphi_j (\xg), \forall j$, where $\Delta$ is a reference value for $|\mdel|$ and $0 \le \varphi_j (\xg) \le 1, \forall j, \xg$:
\begin{eqnarray}
f(\xg)&=&  \fm_{\mdel} (\xg) - \frac{1}{2} \Delta^2 \sum_{j=1}^{3} a_j \varphi^2_j (\xg) \frac{\partial^2 f}{\partial x_j^2} (\xg)
\nonumber\\
&& + O(\Delta^4),
\label{anis-s1}
\end{eqnarray}
where
$$ a_j = \frac{1}{V_{\1}} \int_{V_{\1}}  \xi^2_j  d\xig $$
and
\begin{eqnarray}
\fm_{\mdel \pm {\displaystyle \delta_j \eg_j}}&=&\langle \fm_{\mdel} (\xg)\rangle_{\mdel \pm {\displaystyle \delta_j \eg_j}}
\nonumber\\
&& - \frac{\Delta^2}{2} \sum_{j=1}^{3} a_j \langle  \varphi^2_j (\xg) \frac{\partial^2 f}{\partial x_j^2} \rangle_{\mdel \pm {\displaystyle \delta_j \eg_j}}+O(\Delta^4)\nonumber
\\
&=&\langle \fm_{\mdel} (\xg)\rangle_{\mdel \pm {\displaystyle \delta_j \eg_j}} -
\nonumber\\
&& \frac{ \Delta^2}{2} \sum_{j=1}^{3} a_j  \varphi^2_j (\xg) \frac{\partial^2 f}{\partial x_j^2} + O(\Delta^4),
\label{anis-s2}
\end{eqnarray}
\noindent since from (\ref{anis-s1}) $ \langle  g \rangle_{\mdel} = g + O(\Delta^2), \; \forall g$.

The basic approximation for the anisotropic  non commutation term $\mcalC'_i$ is derived through these expansions as
\begin{eqnarray} 
\tilde{\mcalC}'_{i}(\fm_{\mdel})&=& 
- \sum_{j=1}^{3} \frac{\partial \delta_j}{\partial x_i}
\frac{1}{2\delta_j}
\left(\langle \langle f\rangle_{\mdel}\rangle_{\mdel + {\displaystyle \delta_j \eg_j}}-\langle \langle f\rangle_{\mdel}\rangle_{\mdel - {\displaystyle \delta_j \eg_j}} \right)
\nonumber\\
&&\label{anis-appcomma}\\
 &=& - \sum_{j=1}^{3} \frac{\partial \varphi_j}{\partial x_i}
\frac{1}{2\varphi_j}
\left(\langle \langle f\rangle_{\mdel}\rangle_{\mdel + {\displaystyle \delta_j \eg_j}}-\langle \langle f\rangle_{\mdel}\rangle_{\mdel - {\displaystyle \delta_j \eg_j}} \right),
\nonumber\\
&&
\label{anis-appcommb}
\end{eqnarray}
\noindent while the accuracy of the anisotropic first derivative commutation error and its approximation can be verified to be
\begin{eqnarray} 
{\mcalC}'_{i}(\fm_{\mdel})&=& 
\Delta^2 \sum_{j=1}^{3} a_j \varphi_j \frac{\partial \varphi_j}{\partial x_i}
\frac{\partial^2 f}{\partial x_i^2} + O(\Delta^4)
\nonumber\\
&&\\
\tilde{{\mcalC}'_{i}}(\fm_{\deltag})&=& {\mcalC}'_{i}(\fm_{\deltag})+ O(\Delta^4).
\nonumber\\
&&
\label{anis-accuracy}
\end{eqnarray}
By filtering only in the $j$ direction, expansion (\ref{anis-s1}) becomes
$$ \langle \fm_{\deltag}\rangle_{2\delta_j \eg_j}\rangle = f(\xg) + 2\Delta^2\varphi_j^2(\xg)\frac{\partial^2 f}{\partial x_j^2}(\xg) + O(\Delta^4),$$
which implies
$$
\langle \langle f\rangle_{\deltag}\rangle_{{\displaystyle 2\delta_j \eg_j}}-\langle f\rangle_{\deltag} =
\langle \langle f\rangle_{\deltag}\rangle_{\deltag + \displaystyle{\delta_j \eg_j}}-\langle \fm_{\deltag}\rangle_{\deltag - \displaystyle{\delta_j \eg_j}} + O(\Delta^4),
$$
and,  though keeping the same order of accuracy, the equivalent representation for both relations (\ref{anis-appcomma}, \ref{anis-appcommb}) 
becomes
\begin{eqnarray} 
\tilde{\mcalC}'_{i}(\fm_{\mdel})&=& 
- \sum_{j=1}^{3} \frac{\partial \delta_j}{\partial x_i}
\frac{1}{2\delta_j}
\left(\langle \langle f\rangle_{\mdel}\rangle_{{\displaystyle 2\delta_j \eg_j}}-\langle f\rangle_{\mdel} \right)
\nonumber\\
&&
\label{anis-appcomma2}\\
 &=& - \sum_{j=1}^{3} \frac{\partial \varphi_j}{\partial x_i}
\frac{1}{2\varphi_j}
\left(\langle \langle f\rangle_{\mdel}\rangle_{{\displaystyle 2\delta_j \eg_j}}-\langle f\rangle_{\mdel} \right).
\nonumber\\
&&
\label{anis-appcommb2}
\end{eqnarray}
For flow fields  where the domain grid needs to be stretched along only one direction, say $y$ and whose typical examples are two dimensional wall-bounded flows, the last representation yields very simple approximation formulae.
By adopting the widely used notation $\mdel (\xg) = (\Delta x, \varphi(y) \Delta y, \Delta z)$ in such a case with constants $\Delta x, \Delta y$ and $ \Delta z$, the anisotropic approximation for the first derivative non commutation term results to be
\begin{equation} 
\tilde{\mcalC}'_{y}(\fm_{\mdel})= -
\frac{\partial \varphi}{\partial y}
\frac{1}{2\varphi(y)}
\left(\langle \langle f\rangle_{\mdel}\rangle_{{\displaystyle 2 \varphi(y) \Delta y}}-\langle f\rangle_{\mdel} \right).
\label{anis-wall-bounded}
\end{equation}
The anisotropic non commutation term of the second derivatives, being defined by
\begin{equation}
\mcalC''_{ii}(\fm_{\mdel})= \langle \frac{\partial_2 f}{\partial x_i^2}\rangle_{\mdel} - \frac{\partial^2}{\partial x_i^2} \fm_{\mdel},
\label{anis-defin2}
\end{equation}
can be obtained as in Sec. \ref{par.non-comm} [see (\ref{defin2})--(\ref{commu2})]
\begin{eqnarray}
\mcalC''_{ii}(\fm_{\mdel})= &-& \sum_{j=1}^{3} \frac{\partial^2\delta_j}{\partial x_i^2} \frac{\partial}{\partial \delta_j}\fm_{\mdel} \nonumber\\ 
&-& 2 \sum_{j=1}^{3} \frac{\partial\delta_j}{\partial x_i}(\frac{\partial^2}{\partial\delta_j \partial x_i}\fm_{\mdel}) \nonumber\\ 
&-& \sum_{j,k=1}^{3} \left(\frac{\partial\delta_j}{\partial x_i} \frac{\partial\delta_k}{\partial x_i}\right) \frac{\partial^2}{\partial \delta_j \partial \delta_k}\fm_{\mdel}.
\label{anis-commu2}
\end{eqnarray}
\noindent
Before deriving the approximated form for (\ref{anis-commu2}), while wishing to mantain its fourth order of accuracy,  it is useful to observe that
the terms on the r.h.s. which contain  ${\displaystyle \frac{\partial^2}{\partial \delta_j \partial \delta_k}\fm_{\mdel}}$, $k \ne j$, by (\ref{anis-s1}), are of the same order as the remainder term. Therefore, they do not enter the approximation, which is
\begin{eqnarray}
\tilde{\mcalC}''_{ii}(\fm_{\mdel}) &=& - \sum_{j=1}^{3}
\frac{\delta_j\partial^2_{i}\delta_j+(\partial_j\delta_j)^2}{2\delta_j^2} \left[ \langle \fm_{\mdel}\rangle_{\displaystyle 2\delta_j \eg_j}-\fm_{\mdel}\right] \nonumber\\  
&-&  \sum_{j=1}^{3} \frac{\partial_i\delta_j}{\delta_j} \left[ \langle \partial_i\fm_{\mdel}\rangle_{\displaystyle 2\delta_j \eg_j}-\partial_i\fm_{\mdel} \right]
\label{anis-appr2}
\end{eqnarray}
Again, in analogy with what has been done for the first derivatives, Eq. (\ref{anis-wall-bounded}), the following is obtained
\begin{eqnarray}
\tilde{\mcalC}''_{yy}(\fm_{\mdel}) &=& - 
\frac{\varphi\partial^2_{y}\varphi + (\partial_{y}\varphi)^2}{2\varphi^2(y)} \left[ \langle \fm_{\mdel}\rangle_{\displaystyle 2\varphi(y) \Delta y}-\fm_{\mdel}\right] \nonumber\\  
&-&   \frac{\partial_y\varphi}{\varphi(y)} \left[ \langle \partial_y\fm_{\mdel}\rangle_{\displaystyle 2\varphi(y) \Delta y}-\partial_y\fm_{\mdel} \right]
\label{anis-wall-bounded2}
\end{eqnarray}


\begin{thebibliography}{99}


\bibitem{gm95}
{Ghosal~S., Moin~P.}
{The basic equations for large eddy simulation of turbulent flows in
complex geometries.} {\it J.Comp.Phys.} {\bf 118}, 24--37 (1995).


\bibitem{vdv95}
{Van der Ven~H.}
{A family of large eddy simulation (LES) filters with non uniform filter widths.} {\it Phys. Fluids}, {\bf 7}(5), 1171--1172 (1995).

\bibitem{ft97}
{Fureby~C., Tabor~G.}
{Mathematical and physical constraints on large-eddy simulations} {\it Theoretical and Computational Fluid Dynamics} {\bf 9}(2), 85--102 (1997).


\bibitem{vlm98}
{Vasilyev~O.V., Lund~T.S. and Moin~P.}
{A general class of commutative filters for LES in Complex Geometries.} {\it J.Comp.Phys.} {\bf 146}, 82--104 (1998).

\bibitem{ghosal99}
{Ghosal~S.}
{Mathematical and physical constraints on large-eddy simulation of turbulence.}
{\it AIAA Journal}, {\bf 34}(4), 425--433 (1999).

\bibitem{s01}
{Sagaut~P.}
{\it Large eddy simulation for incompressible flows}, Springer Verlag, Berlin-Heidelberg  (2001).



%







\bibitem{g91}
{Germano~M.,Piomelli~U.,Moin~P. and Cabot~W.H.}
{A dynamic subdrid-scale eddy viscosity model.} 
{\it Phys. Fluids A} {\bf 3}(7), 1760--1765 (1991).

\bibitem{g92}
{Germano~M.}
{Turbulence, the filtering approach.} 
{\it J.~Fluid Mech.} {\bf 236}, 325--336 (1992).

\bibitem{bfr80}
{Bardina J., Ferziger J.H., Reynolds W.C.}
{Improved subgrid models for large eddy simulation.}
{\it AIAA Paper} {\bf 80}--1357 (1980).


\bibitem{sma}
{Smagorinsky J.}
{General circulation experiments with the primitive equations.} 
{\em Monthly Weather Review}, {\bf 91}, 99--164 (1963).

\bibitem{passoni1}
G.Alfonsi, G.Passoni, L.Pancaldo and D. Zampaglione
A spectral-finite difference solution of the Navier-Stokes equations in three dimensions,
{\it Int. J. Numer. Meth. Fluids}, {\bf 28}, 129--142 (1998).

\bibitem{passoni2}
G.Passoni, G. Alfonsi, G.Tula and U.Cardu
A wavenumber parallel computational code for the numerical integration of the Navier-Stokes equations,
{\it Parallel Computing}, {\bf 25}, 593--661 (1999).

\bibitem{mkm1999}
Moser R.D., Kim J., Mansour N.N.,
Direct numerical simulation of turbulent channel flow up to $Re_\tau=590$,
{\it Phys. Fluids}, {\bf 11}(4), 943--945 (1999).


\bibitem{mk82}
Moin P. and Kim J.
{Numerical investigation of turbulent channel flow},
{\em J. Fluid Mech.}, {\bf 118}, 341--377 (1982).


\bibitem{d70}
Deardorff J.W.
A numerical study of three-dimensional turbulent channel flow at large Reynolds numbers,
{\it J. Fluid Mech.}, {\bf 41}, 435--452 (1970).

\bibitem{s75}
Schumann U.
{Subgrid scale model for finite difference simulations of turbulent flows
in plane channels and anuli.}
{\it J.Comput.Phys.}, {\bf 18}, 376--404 (1975).

\bibitem{g87}
Gr\"otzbach G., in {\it Encyclopedia of Fluid Mechanics}, vol.6, ed. N.P. Cheresimov, Gulf West Orange, New York (1987).

\bibitem{pfm89}
{Piomelli~U., Ferziger~J. \& Moin~P.}
{New approximate boundary conditions for large eddy simulations of wall-bounded flows.}
{\it The Physics of Fluids A} {\bf 1}(6), 1061--1068 (1989).

\bibitem{h90}
Horiuti, K.
{Higher order terms in the anisotropic  representation of Reynolds stresses},
{\em Phys. Fluids A}, {\bf 2}(19), 1708--1710 (1990).


\bibitem{cc96}
Carati D. and Cabot W.
{Anisotropic eddy viscosity models. Proceeding of the summer program - Center for Turbulence Research, Stanford}, 249--259 (1996).


\bibitem{it02}
Iovieno M., Tordella D.
The angular momentum equation for a finite element of fluid: a new representation and  application to turbulent flows,
{\it Phys. Fluids}, {\bf 14}(8), 2673--2682 (2002).

\bibitem{ghosal96}
Ghosal S.
{An analysis of numerical errors in Large-Eddy Simulations of Turbulence}
{\it J. Comput.  Phys.}, {\bf 125}, 187--206 (1996).
















\end{thebibliography}
\end{document}